\documentclass[aps,a4paper]{revtex4}
\usepackage{epsfig}
\usepackage{graphicx}
\usepackage{amsmath,amssymb,color}
\usepackage[english]{babel}

\parskip=\medskipamount

\def \beq {\begin{equation}}
\def \eeq {\end{equation}}
\def \beqn {\begin{eqnarray}}
\def \eeqn {\end{eqnarray}}
\def \la {\left \langle}
\def \ra {\right \rangle}

\def \lam {\lambda}
\def \Tr {{\rm Tr}}
\def \dim {{\mathrm{dim}}}

\def \pr {\partial}

\def \th {\theta}
\def \disp {\displaystyle}

\def \eps {\varepsilon}

\newcommand{\eq}[1]{(\ref{#1})}
\newcommand{\fig}[1]{Fig.\ref{#1}}

\begin{document}

\title{Two faces of Douglas-Kazakov transition: from Yang-Mills theory to random walks and beyond}

\author{Alexander Gorsky$^{1,2}$, Alexey Milekhin$^{1,3,4}$, and Sergei Nechaev$^{5,6}$}

\affiliation{$^1$Institute of Information Transition Problems, B.Karetnyi 19, 127051 Moscow, Russian Federation \\ $^2$Moscow Institute of Physics and Technology, Dolgoprudny 141700, Russian Federation \\ $^3$Princeton University, Physics Department, 08544 NJ, USA \\ $^4$Institute for Theoretical and Experimental Physics, B. Cheryomushkinskaya 25, 117218 Moscow, Russian Federation \\ $^5$Interdisciplinary Scientific Center Poncelet (CNRS-UMI 2615), Bolshoy Vlasyevskiy 11, 119002 Moscow, Russian Federation \\ $^6$P.N.Lebedev Physical Institute, RAS, 119991 Moscow, Russian Federation}

\begin{abstract}

Being inspired by the connection between 2D Yang-Mills (YM) theory and (1+1)D "vicious walks" (VW), we consider different incarnations of large-$N$ Douglas-Kazakov (DK) phase transition in gauge field theories and stochastic processes focusing at possible physical interpretations. We generalize the connection between YM and VW, study the influence of initial and final distributions of walkers on the DK phase transition, and describe the effect of the $\theta$-term in corresponding stochastic processes. We consider the Jack stochastic process involving Calogero-type interaction between walkers and investigate the dependence of DK transition point on a coupling constant. Relying on the relation between large-$N$ 2D $q$-YM and extremal black hole (BH) with large-$N$ magnetic charge, we speculate about a physical interpretation of a DK phase transitions in a 4D extremal charged BH.
\end{abstract}

\maketitle

\tableofcontents

\section{Introduction}
\label{sect:intro}

The third order phase transitions, known also as Douglas-Kazakov (DK) transitions, were
comprehensively studied in \cite{dk} in the large-$N$ two-dimensional Yang-Mills
(YM) theory on a sphere. It was understood that the DK transition is an reincarnation of a
Gross-Witten-Wadia phase transition \cite{gw,wadia} in the lattice version of gauge theories.
In the matrix model framework, the DK transition results in changing the density eigenvalue
support: below the transition point, the density enjoys the one-cut solution, while above the
transition it has two-cut form corresponding to an elliptic curve. Within the transition region,
the eigenvalue density is governed by the universal Tracy-Widom distribution. In terms of the 2D Yang-Mills (YM) theory on a sphere $S^2$, the phase transition occurs at some critical value of the sphere radius, $R_c$. It has been found that the YM partition function is dominated by the zero-instanton charge sector in a weak coupling regime (below the transition point), while at strong couplings (above the transition), the instantons dominate \cite{grm}. Similar phase transition was found also for the Wilson loop in the 2D YM theory \cite{olesen}.

Phase transitions of the same class occur in a 2D YM theory on the cylinder and on the disc \cite{grm}. In these cases, the phase transition points strongly depend on the boundary holonomies. Quite recently, the generalizations of the 2D YM to the 2D $q$-YM \cite{jafferis,marino05,qym} and to the 2D $(q,t)$-YM \cite{szabo13,aganagic12} were developed, and a very rich structure of the DK-like phase transition has been found. In these cases, the transition points again separate different phases: the "perturbative" (weak coupling) phase and the "non-perturbative" (strong
coupling, or instanton-dominated) one. However, contrary to the ordinary DK transition, for the $q$-YM one sees multiple transitions driven by different types of instantons. Below we exploit close relations among three seemingly different systems: i) deformed large-$N$ topological field theories in 2D and 3D space-time \cite{deharo1, deharo2,deharo3}, ii) directed (1+1)D vicious random walks (VW) with different initial and boundary conditions \cite{fms,fms2}, and iii) entropy of extremal (i.e zero-temperature) charged black holes (BH) in 4D at large magnetic charge \cite{vafa,osv,ooguri}.

The DK critical behavior can be traced not only for the partition function but for other important variables. The hydrodynamic viewpoint on a real-time stochastic dynamics of $N$ (1+1)D vicious walkers deals with the problem of computation the fermionic multi-point correlation function $\rho_N(x_1,x_2,...x_k,t)$ for $1\le k\le N$. A particularly interesting object is the resolvent in the YM theory defining the one-particle correlation function (the density). It obeys the complex-valued Burgers-Hopf equation \cite{grm} with specific boundary conditions. We emphasize below that the hydrodynamic point of view becomes very useful in two aspects. On one hand, it enables us to consider a fermionic system with large number of particles. That problem itself has a lot in common with the overlapping of wave functions in a multi-fermionic model studied in \cite{abanov1,abanov2}. On the other hand, even the one-point correlation function, $\rho_N(x,t)$, which satisfies the Burgers-Hopf equation, has some far-reaching connections. For example, $\rho_N(x,t)$ in the large-$N$ vicious walks ensemble, shares limiting properties of the statistics of (1+1)D Dyck excursions with fixed area below the curve. The last problem is known to be connected with the generating function of algebraic invariants (Jones, HOMFLY) of some specific series of torus knots \cite{bgn}.

The important question concerns the hydrodynamic identification and interpretation of DK transition. The typical example of the critical behavior can be interpreted as an overturn of the solution at finite times due to the non-linearity (as in a Burgers equation at small viscosity). However, as shown in \cite{grm, nowak1}, the criticality pattern is different. For the initial state of the fluid prepared in a form of a local bump, the transition happens when two fronts of a bump moving in opposite directions on the circle, collide at some finite time. At the transition point the fluid density closes gaps on the circle. Such a behavior means that one deals with the quantum hydrodynamics, and the solution upon collision corresponds to the strong coupling phase dominated by instantons in YM setting and nontrivial windings in VW setting.

From the standpoint of the DK transition, the YM--VW correspondence is generalized in our paper
along the following lines:

\begin{itemize}
\item We demonstrate that the DK phase transition in the 2D YM on the cylinder and the disc with fixed holonomies amounts to the phase transition in the stochastic process with prescribed configurations of extremities (initial and final points);

\item We take into account the $\theta$-term in the 2D YM, fixing the total instanton number, and discuss the vicious walks counterpart of corresponding theory where we introduce the chemical potential for windings on a circle;

\item We add a Wilson line in a particular representation to the 2D YM on the cylinder which yields the trigonometric Calogero model \cite{gn1,gn2}. From the VW viewpoint, it involves the modified interaction between walkers, and corresponds to the so-called "Jack stochastic process".
\end{itemize}

The problems related to the DK phase transition can be reformulated in terms of the string theory, thus uncovering the connection with black holes. It has been found in \cite{vafa}, that the 2D YM theory on a torus can be realized as the worldvolume theory on $N$ D4-branes wrapping the $O(-p)\rightarrow T^2$ subspace inside $O(-p)\times O(p)\rightarrow T^2$ Calabi-Yau manifold. On the other hand, $N$ such D4-branes yield the description of the 4D black hole with the magnetic charge $N$. Explicitly, the relation between partition functions of the 4D black hole, $Z^{4D}_{BH}$, the 2D YM, $Z^{2D}_{YM}$, and the topological string, $Z_{top}$, schematically is as follows \cite{osv}
\beq
Z^{4D}_{BH}=Z^{2D}_{YM}= |Z_{top}|^2
\eeq
The correspondence between the 4D charged black hole and the 2D YM was generalized in \cite{ooguri} to the arbitrary genus $g$ of the base in Calabi-Yau. However, it turned out that the $q$-deformed 2D YM has to be considered as the $g\neq 1$ theory, where the additional integration over the holonomies is involved. Similar relation between the partition function of the $(q,t)$ 2D YM and of the refined black hole indices was discussed in \cite{aganagic12}. The BH is extremal and to some extend can be regarded as the particle with a huge degeneracy, corresponding to multiplicity of the D2-D0 branes on $N$ D4 branes forming the extremal charged black hole.

Using the maping between $q$-deformed YM in 2D and the black hole partition function \cite{vafa,
ooguri}, we identify values of the black hole chemical potentials corresponding to the DK phase
transition. Since we consider the limit of BH with large magnetic charge, the very phenomena and the configurations specific for this limit, are of special interest for us. It was argued in
\cite{bolognesi} that the configuration with a large magnetic charge tends to form a spherical
shell monopole which, at some critical value of parameters, gets transformed into the magnetically
charged BH. This is the magnetic counterpart of the phenomena known for electrically charged BH
which undergoes the superconducting phase transition, and BH with the fermionic environment where the electronic star gets formed at the point of a 3rd order phase transition (see \cite{hartnoll} for the review). We conjecture that the DK phase transition at the BH side corresponds to the transition between the "monopole wall" and "magnetically charged BH" discussed in \cite{bolognesi}.

The relation between the DK transition and BH physics has been discussed in \cite{alvarez,wadia07}
from a different perspective. In these papers the large-$N$ 4D gauge theory on the manifold involving $S^3$ is considered  holographically. There is some compact direction treated as the thermal circle, and the gravity background involving the BH. In such a setup one can identify a Page-Hawking transition, which is a version of a "small BH"--string transition suggested long time ago in \cite{susskind, polchinski}. The interpretation of DK transition as a "small BH"--string transition was suggested in \cite{alvarez}, however the parameters of the gauge theory were not related to the BH charges and chemical potentials. Our picture differs from this scenario.

The paper is organized as follows. In Section II we recall the key points concerning the DK phase
transition in the 2D YM theory and show that the YM--VW correspondence yields the new critical behavior at the VW side for various specific initial states. In Section III we consider the duality between Calogero-type integrable systems and discuss the emergence of the DK transition. Section IV is devoted to the hydrodynamic aspects of the DK transition and its interpretation as an "ortogonality catastrophe". In Section V we discuss the $q$-YM--BH correspondence from the DK phase transition viewpoint. The results and open questions are summarized in Discussion.

\section{Douglas--Kazakov phase transitions in the 2D Yang-Mills theory}
\label{sect:dk}

\subsection{2D Yang-Mills on a sphere $S^2$}

Let us briefly review the original setting of Douglas-Kazakov (DK) phase transition developed in
\cite{dk} for a Yang-Mills (YM) theory on a sphere $S^2$. The YM partition function on a
two-dimensional surface $\Sigma_g$ of genus $g$ and area $A$ reads
\beq
Z_g(g_{YM},A) = \int \left[d\mathbf{A}\right] \left[d\mathbf{\Phi}\right]
\exp\left(\int_{\Sigma_g} d\sigma dt \left(\Tr\,\mathbf{\Phi} \mathbf{F} + \frac{g_{YM}^2}{N}
\Tr\,\mathbf{\Phi}^2\right)\right).
\label{eq:1}
\eeq
where $\mathbf{A}$ is the vector potential of the gauge field $\mathbf{F}=d \mathbf{A}$, $g_{YM}$ is the 't Hooft coupling constant, and $\mathbf{\Phi}$ is the adjoint scalar. As it is explicitly shown in \cite{dk}, the partition function (\ref{eq:1}) for the $U(N)$ gauge group can be rewritten as a sum over representations:
\beq
Z_g(g_{YM},A)=\sum_R (\dim\, R)^{2-2g}e^{-\frac{Ag_{YM}^2}{2N}C_2(R)},
\label{eq:part_rep}
\eeq
Different representations $R$ are labelled by Young diagrams, i.e. by sets of $N$ ordered non-negative integers $\left\{n_1\ge n_2\ge \dots \ge n_N\right\}$. The functions $C_2(R)$ and $\dim\, R$ are correspondingly the quadratic Casimir (Laplace operator) and the dimension of $R$:
\beq
C_2(R)=\sum_{i=1}^N n_i\left(n_i-2i+N+1 \right), \qquad \dim\, R=\prod_{i>j}\left(
1-\frac{n_i-n_j}{i-j} \right).
\label{eq:rep_def}
\eeq
In what follows we mostly are interested in the genus $g=0$ case, corresponding to the topology of an $S^2$ sphere. In the large-$N$ limit keeping the combination $g_{YM}^2 N$ fixed, the system experiences a third-order phase transition at some critical value of $A$. In terms of the continuous variable $h(x)$, describing the shape of the Young diagram,
\beq
h(x)=-\frac{1}{2}+\frac{i-n_i}{N}, \qquad x=\frac{i}{N}
\label{eq:cont}
\eeq
the effective YM action for $N\gg 1$ can be written as:
\beq
S_{eff}[h(x)]=-\int_0^1\int_0^1 dx dx'\,\log \left|h(x)-h(x')\right|+\frac{A g_{YM}^2}{2}\int_0^1 dx\, h^2(x)-\frac{A g_{YM}^2}{24}.
\label{eq:action_cont}
\eeq
For a Young diagram, we have, by definition, a constraint on its shape, $h(x)$, meaning that $n_i$ should sequentially increase with $i$. This requirement in the limit $N\gg 1$ gets mapped onto the condition
\beq
\rho(h)\le 1
\label{eq:young}
\eeq
where $\rho(h)=\frac{\partial x(h)}{\partial h}$.

The phase transition occurs, when the classical solution of (\ref{eq:action_cont}) ceases to
satisfy the constraint $\rho(h)\le 1$. It reads as
\beq
\frac{\delta S_{\rm eff}[h(x)]}{\delta h(x)}=0 \quad \Rightarrow \quad \rho=\frac{Ag_{YM}^2}{2\pi} \sqrt{\frac{4}{Ag_{YM}^2}-h^2}.
\label{eq:saddle_point}
\eeq
hence, the DK phase transition occurs at
\beq
A\equiv A_{cr}=\frac{\pi^2}{g_{YM}^2}.
\label{eq:crit}
\eeq

\begin{figure}[ht]
\epsfig{file=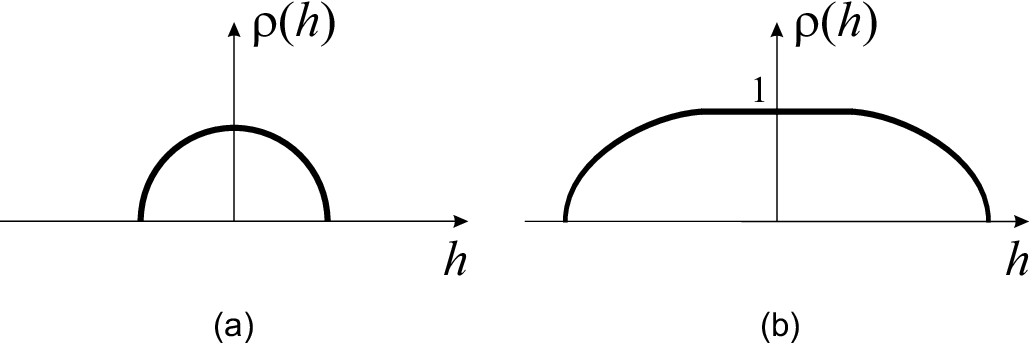, width=12cm}
\caption{Density profile: (a) below the phase transition, (b) above the phase transition.}
\label{fermi}
\end{figure}
Beyond $A_{cr}$, the density $\rho(h)$ of eigenvalues develops a plateau around $h=0$ -- see \fig{fermi}.

The transition has a remarkable property: it is of the third-order. Near the transitions the free energy behaves as:
\beq
\Delta F \sim \frac{g_{YM}^6}{2\pi^6} (A-A_{cr})^3
\eeq

\subsection{2D Yang-Mills on a cylinder and a disc: account for boundary holonomies}

Here we summarize the results for the 2D YM in other underlying geometries \cite{grm}. On a disc and on a cylinder the transition amplitudes depend on the boundary holonomies (one for a disc and two for a cylinder). In both cases the DK phase transition happens at particular values of the area of a disc or a cylinder. In several cases one can derive explicit equations for the critical value $A_{cr}$ at which the phase transition occurs.

From now on we will also include 2d $\theta$-term in the Yang--Mills theory. Defining the non-trivial holonomies, $U_1$ and $U_2$ around the boundaries of the cylinder, the partition function on the cylinder, $Z_{cyl}$, can be written as:
\beq
Z_{cyl}(g_{YM}^2, A |U_1, U_2)=\sum_R \chi_R(U_1)\chi_R(U_2^\dagger)\, e^{-\frac{Ag_{YM}^2}{2N} C_2(R)+i \theta |R|},
\label{eq:partition_cylinder}
\eeq
where $|R|$ is the total number of boxes in the Young diagram and $\chi_R$ are characters (Schur polynomials) of the representation $R=\left\{ n_1, \dots, n_N \right\}$.
\beq
\chi_R(U)=\frac{\det_{i,j} (e^{i(n_i-i)\theta_j})}{\prod_{i<j}\left(e^{i\theta_i}-e^{i\theta_j}
\right)},
\label{eq:character}
\eeq
and $0\le\theta_i\le 2\pi$ are phases of the eigenvalues of the unitary matrix $U$, below we will refer to them as eigenvalues for brevity.

For the uniform distribution, $\sigma(\theta)$, of the eigenvalues of the boundary holonomies on the interval of length $c$ on the cylinder
\beq
\sigma(\theta)=\left\{
\begin{array}{ll}
\frac{1}{2c}, &|\theta|\le c, \medskip \\ 0, & \mbox{otherwise},
\end{array} \right.
\label{eq:distrib}
\eeq
the phase transition occurs at the value $A_{cr}$ determined by the equation
\beq
\tanh^2\frac{\pi c}{A_{cr}} =\tanh \frac{c^2}{A_{cr}}.
\label{eq:cyl_crit}
\eeq
Taking $c=\frac{ig_{YM}N}{2}$, in the large-$N$ limit of (\ref{eq:q_cyl}) we arrive at a condition for the phase transition in $q$-YM \cite{marino05}:
\beq
e^{\frac{g_{YM}^2N^2}{A_{cr}}}=1+\tan^2\frac{\pi g_{YM}N}{A_{cr}}.
\label{eq:dk_qym}
\eeq

If one boundary shrinks to the point, as shown in the \fig{fig:dk_f01}, we arrive at the
two-dimensional YM on a disc, where one common extremity (i.e. initial or final point) of the bunch
of walkers is located at the center of the disc. The DK phase transition occurs in this case at the
critical disc area defined by the condition
\beq
\frac{\pi}{A_{cr}}= \int \frac{\sigma(s)\,ds}{\pi - s}
\eeq
where $\sigma$ has support on a interval $[-c,c] \subset [-\pi,\pi]$ (see \cite{grm} for details).

\subsection{Phase transition in $\beta$-ensemble}

After reviewing known results about standard Yang--Mills theory, let us consider a $\beta$-ensemble. The DK phase transition can be straightforwardly extended to the case of generic $\beta$-ensemble. Consider the partition function
\beq
Z\left(\beta,g_{YM},A \right)= \sum_{\left\{n_i\right\}} \Delta(\mathbf{n}-\beta
\mathbf{i})^{2\beta}\exp \left(-\frac{A g_{YM}^2}{2}\sum_i{n_i(n_i - 2 \beta i + N+1)}\right)
\eeq
where $\Delta(\mathbf{n}-\beta \mathbf{i})$ is the Vandermonde determinant,
\beq
\Delta(\mathbf{n}-\beta \mathbf{i})=\prod_{i<j}(n_i-n_j-\beta (i-j))
\label{vand1}
\eeq
Eq.\eq{vand1} is the generalization of \eq{eq:rep_def}: at $\beta=1$ we return to \eq{eq:part_rep}--\eq{eq:rep_def}.

Since we are interested in the large-$N$ limit, it is convenient to pass, as in \eq{eq:cont}, to the continuum variable $h(x)$:
\beq
h(x)=-\frac{1}{2}+\frac{\beta i-n_i}{N}, \qquad x=\frac{i}{N}
\label{eq:cont_beta}
\eeq
Thus, instead of (\ref{eq:young}), we get a modified condition:
\beq
\rho(h) \leq \frac{1}{\beta}
\label{eq:young2}
\eeq
and the third-order phase transition occurs at
\beq
A\equiv A_{cr}=\frac{\pi^2}{\beta g_{YM}^2}.
\eeq

\section{Douglas-Kazakov transition for free fermions, directed vicious walks and orthogonality catastrophe}

It is known \cite{fms,fms2} that the partition function of a 2D Yang-Mills theory on a sphere $S^2$ with the $U(N)$ gauge group coincides with the partition function of a bunch of $N$ directed one-dimensional non-intersecting (vicious) walks in a $(t,x)$-plane. Each path is stretched along the $t$-axis and has a length $t$, while along $x$-axis the system is periodic on a strip width $L$. The parameters $t$ and $L$ are absorbed in the YM coupling constant, $g_{YM}$. Changing the gauge group $U(N)$ to $Sp(2N)$, or to $SO(2N)$, one gets the partition functions of vicious walks in the strip of width $L$ with the Dirichlet (for $Sp(2N)$), or Neumann (for $SO(2N)$) boundary conditions at $x=0$ and $x=L$.

In a real-time stochastic Brownian dynamics, the DK transition is identified as follows. Consider $N$ ($N\gg 1$) vicious (1+1)D Brownian walkers (or free fermions) on a circle, and compute the probability for them to reach some prescribed final state at time $T$ starting from some fixed initial configuration. This probability obeys the multiparticle diffusion equation similar to the Schr\"odinger equation in the Euclidean time. Replacing the Laplace operator by another self-adjoint one, we equivalently describe the system of fermionic particles by the Fokker-Planck equation for "Dyson random waks", which in turn, can be derived from the real-time Langevin equation with random forcing and account for condition that trajectories do not intersect each other (obey the Pauli principle). The properly normalized partition function of the large-$N$ 2D YM theory on the sphere, turns out to coincide with the reunion probability of $N$ "Brownian fermions" (Schur process) on the circle upon the proper identification of parameters \cite{deharo1,fms}. The reunion probability undergoes a 3rd order phase transition at some critical $T_c$ and has a lot in common with the "orthogonality catastrophe" for the many-body fermionic system \cite{abanov2}.

The interpretation of the "phase transition" at some finite time $T_c$ in the simplest case when initial and final states of all fermions are placed in the vicinity of the same point on the circle, is as follows. At $T<T_c$ the fermionic trajectories do not wind around the circle while at $T>T_c$ the total winding number during the process becomes finite. Roughly we could say that the DK transition point separates the perturbative and non-perturbative regimes in the evolution for prescribed initial and final states. To some extend, the "orthogonality catastrophe" is resolved non-perturbatively.

\subsection{Short reminder: From (1+1)D vicious walks to 2D YM on $S^2$}
\label{sect:two-sphere}

Here we briefly review the relation between 2D YM on $S^2$ and VW on the circle. The evaluation of the 2D YM partition function on the sphere with the $U(N)$ gauge group can be reformulated in terms of the computation of the reunion probability of $N$ vicious walkers \cite{fms,fms2,deharo1, deharo2,deharo3}. Consider Brownian dynamics of $N$ repulsive particles with the fermionic statistics on a periodic one-dimensional lattice with the period $L$ (i.e on a circle of circumference $L$), starting from the initial distribution, $\mathbf{x}=\{x_1,..., x_N\}$, and arriving after time $t$ at a final distribution, $\mathbf{y}=\{y_1,...,y_N\}$. The reunion probability, $P(t,\mathbf{x},\mathbf{y})$, can be constructed from the eigenfunction, $\Psi_N(\mathbf{p}|\mathbf{x})$, as the fully antisymmetric Slater determinant (known equivalently as Fisher determinant, Karlin-McGregor, or Lindstrom-Gessel-Viennot formulae) constructed on the basis of wave functions $\psi(p_j|x_k)$ of individual noninteracting particles:
\beq
\Psi_N\left(\mathbf{p} | \mathbf{x}  \right)=\frac{1}{\sqrt{N!}}\det\psi(p_j|x_k), \quad
\psi(p_j|x_k)=e^{i p_j x_k} \qquad \qquad 1\le j,k\le N
\label{eq:slater}
\eeq
Here $\mathbf{x}=\left\{x_1,\dots,x_N \right\}$ are the coordinates of the particles and
$\mathbf{p}=(p_1,...,p_N)$ are the corresponding momenta. Consider a configuration when all
particles are localized near a single point and after time $t$ return to the same localized region on a circle. The probability of such a process can be written using (\ref{eq:slater}):
\beq
P_N(t,|\mathbf{x}, \mathbf{y}=\mathbf{x})\Big|_{\mathbf{x}\to 0}=\langle\Psi_N|e^{-t
H_0}|\Psi_N\rangle\Big|_{\mathbf{x}\to 0}=\sum_{{\mathbf p}}\Psi\left(\mathbf{p}|\mathbf{x} \right)
\Psi^*\left(\mathbf{p}| \mathbf{x} \right)e^{-tE(\mathbf{p})}\Big|_{\mathbf{x}\to 0},
\label{eq:reunion1}
\eeq
where each $E(p_j)$ is the eigenvalue of the free Schr\"odinger operator $H_0=-D \sum_{k=1}^N \frac{\partial^2}{\partial x_k^2}$ for the non-stationary wave function, $\psi(t,x)$, on a circle with the circumference $L$,
$$
\left\{\begin{array}{l} \disp \frac{\partial \psi(t,x)}{\partial t} = D\frac{\partial^2
\psi(t,x)}{\partial x^2} \medskip \\ \disp \psi(t,x)=\psi(t,x+L) \end{array} \right.
$$
The corresponding values of $E(p_j)$ are:
\beq
E(\mathbf{p})=\sum_{j=1}^N E(p_j), \qquad E(p_j)=\frac{4\pi^2 D}{L^2} p_j^2, \qquad p_j=0,1,2,...
\label{eq:E}
\eeq

As we have mentioned above, we rely on the fact that the free Schr\"odinger equation in the imaginary time coincides with the Fokker-Plank equation, where the diffusion coefficient $D$ is related to the mass by $D=1/2m$. Taking in (\ref{eq:reunion1}) the limit $\mathbf{x}\to 0$, we arrive at
\beq
P_N\left( t,L| \mathbf{x}\to 0\right)=C\delta^{N(N-1)/2} \sum_{\mathbf{p}\in
Z}\prod_{i<j}^N(p_i-p_j)^2 e^{-\frac{4 \pi^2 Dt}{L^2}\sum\limits_{i=1}^N p_i^2},
\label{eq:reunion2}
\eeq
where $\delta=x_i-x_{i+1} \rightarrow 0$ and $C$ is the normalization factor. For $N\to\infty$ we keep the density of particles $\frac{L}{N}$ finite, and the combination $tD$ scales as $N$, since we keep the total mass $m N$ finite as well. Comparing the reunion probability (\ref{eq:reunion2}) with the partition function (\ref{eq:part_rep}), we see that the reunion probability upon the identification:
\beq
{g_{YM}^2 A}=\frac{8\pi^2DNt}{L^2}
\label{eq:ident}
\eeq
coincides (up to the normalization factor) with the partition function of the Yang-Mills theory on a sphere with the $U(N)$ gauge group. Therefore, the phase transition for free fermions on a circle transition occurs at:
\beq
\frac{8DtN}{L^2} = 1
\label{eq:transit}
\eeq
Note the following subtlety: at first sight in \eq{eq:reunion2} the sum goes over all integer momenta $p_i$. However, since we are dealing with fermions, all momenta should be distinct. It is equivalent to requirement that $p_1< \dots < p_N$.

\subsection{Vicious walkers and 2D YM on the disc and cylinder}

Consider the generalization of the result obtained in previous section and discuss the dependence of the transition point on the distribution of extremities of vicious walkers. Here we consider vicious walks on a periodic lattice along $x$-axis with two different configurations of initial and final points: (i) at both extremities the points of vicious walks ($U(N)$ case) are equally-spaced distributed within some interval $\eps$ along $x$-axis; (ii) initial starting points are equally-spaced distributed within the interval $\eps$, while spacing between terminal points tends to 0.

\begin{figure}[ht]
\epsfig{file=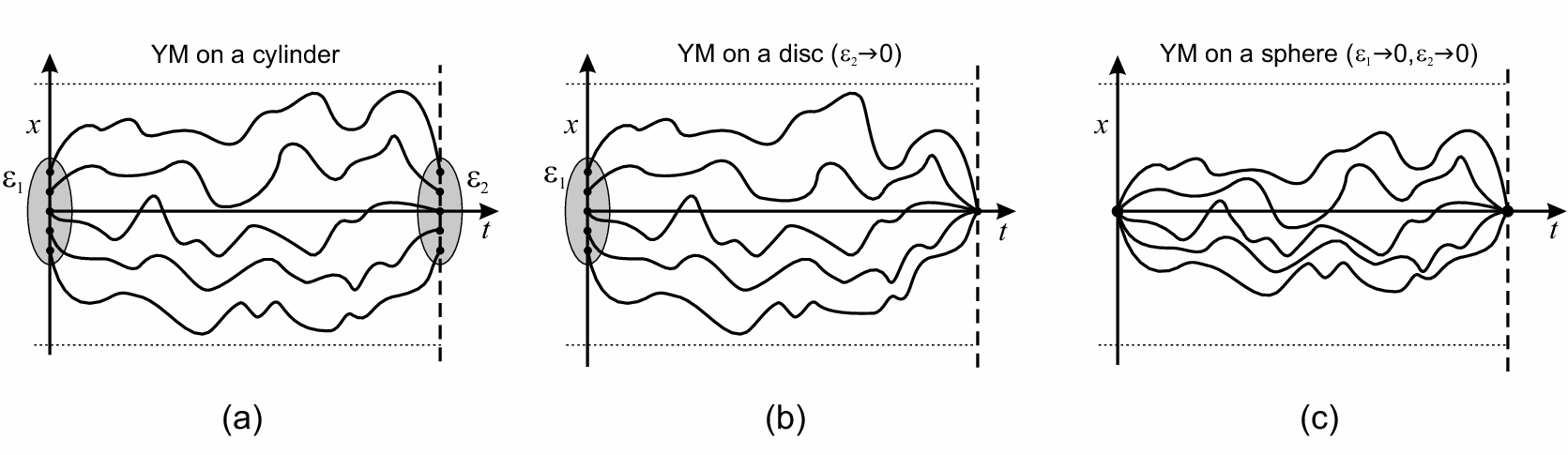, width=17cm}
\caption{The 2D YM on a cylinder (a) is reduced to the 2D YM a disc (b) when left extremities
are kept on finite support ($\eps_2\to 0)$, and right ones are set to zero, and to the 2D YM on a
sphere (c) when $\eps_1\to 0$ and $\eps_2\to 0$.}
\label{fig:dk_f01}
\end{figure}

In the frameworks of the field-theoretic description, the case (i) corresponds to the 2D $U(N)$ YM theory on a \emph{cylinder}, and the case (ii) -- to the 2D $U(N)$ YM theory on a \emph{disc}. We also show that the case (ii) at $\eps\to 0$ is reduced to the 2D $U(N)$ YM on a \emph{sphere}. Schematically these cases are depicted in the \fig{fig:dk_f01}.

Mapping of the 2D YM theory onto vicious walkers on the cylinder goes as follows. The probability
for $N$ vicious walkers with the boundary conditions $(x_1,\dots x_N)$ and $(y_1,\dots y_N)$
corresponds to the transition amplitude in the 2D YM on the cylinder. The initial distribution of
points sets the holonomy of the gauge field at the left boundary of the cylinder, and the final distribution of walkers -- to the holonomy at the right boundary of the cylinder.
This can be easily seen if one rewrites \eq{eq:slater} in terms of Schur functions:
\beq
\Psi_N\left(\mathbf{p} | \mathbf{x}  \right)=\frac{1}{\sqrt{N!}} \chi_R(e^{i x_1},\dots, e^{i x_N})
\Delta(e^{i \mathbf{x}})
\eeq
where we have introduced an auxiliary Young diagram, $R$: $n_i=p_i+i$. We take the energy to be
proportional to the quadratic Casimir of $R$:
\beq
E(\mathbf{p})=\cfrac{1}{2m}\left(\cfrac{2 \pi}{L} \right)^2 \sum_{i=1}^N
n_i(n_i-2i+N+1)=\cfrac{1}{2m}\left(\cfrac{2 \pi}{L} \right)^2 \sum_{i=1}^N \left(
p_i+\frac{N+1}{2} \right)^2 + const
\label{eq:E}
\eeq
Equation \eq{eq:E} actually means that the physical momenta are $p_i+\frac{N+1}{2}$. Therefore, we can rewrite the transition amplitude as:
\begin{multline}
\la \Psi_N(\mathbf{p}|\mathbf{x})|e^{-t H_L}| \Phi_N(\mathbf{q}|\mathbf{x})\ra \Big|_{\mathbf{x}\to
0} = C \Delta(e^{-i \mathbf{p}}) \Delta(e^{i \mathbf{q}}) \sum_R \chi_R(e^{-i p_1},\dots,e^{-i
p_N}) \chi_R(e^{i q_1},\dots,e^{i q_N}) \\ \times \exp\left(-\cfrac{2 \pi^2 t}{m L^2} \sum_{i=1}^N
p_i^2\right)
\label{eq:cyl}
\end{multline}
The expression \eq{eq:cyl} exactly coincides with the partition function of vicious walks on a cylinder -- compare (\ref{eq:cyl}) and (\ref{eq:partition_cylinder}). Thus, the DK phase transition happens at some critical value of the distance between the initial positions of the walkers.

\section{Interacting random walkers: Calogero model and Jack process}

So far we considered the free fermions, which means $\beta=1$ in terms of $\beta$-ensembles. The DK
critical behavior is manifested itself in matrix elements $W(\mathbf{\phi}|\mathbf{\theta})=\la \mathbf{\phi} | e^{-t H_L} | \mathbf{\theta} \ra$ of the evolution operator $H_L$ between some initial and final $N$-fermion states on a circle. For generic $\beta$ the free fermions on a circle get transformed into the particles interacting via the trigonometric Calogero potential. The wave functions and the spectrum of the model is exactly known, and can be conveniently represented in terms of Jack polynomials. Considering the amplitude $W(\mathbf{\phi}|\mathbf{\theta})$, one can find the DK transition for matrix element corresponding to the reunion of world trajectories of particles.

We can transform the Fokker-Plank (FP) equation into the Schr\"odinger equation by the self-adjoint change of variables even in the interacting case. The potential in the FP equation and in the Schr\"odinger equation are related as the ordinary potential $V$ and the "superpotential" $W$:
\beq
V({\bf x})=D \left[ (\nabla W({\bf x}))^2-\Delta W({\bf x}) \right]
\label{eq:super}
\eeq
Indeed, it is easy to see that the Schr\"odinger equation
\beq
\cfrac{\pr \psi({\bf x},t)}{\pr t}=\cfrac{1}{2m} \Delta \psi({\bf x},t) - V({\bf x}) \psi({\bf
x},t)
\label{eq:super2}
\eeq
becomes the FP equation
\beq
\cfrac{\pr P({\bf x},t)}{\pr t}=D \Delta P({\bf x},t) + D \sum_i \cfrac{\pr}{\pr x_i}
\left(\cfrac{\pr}{\pr x_i} W({\bf x}) P({\bf x},t) \right)
\label{eq:super3}
\eeq
upon the substitution $D=\frac{1}{2m}$ and $P({\bf x},t)=\psi({\bf x},t) e^{-W({\bf x})}$.

Recall that the Calogero-Sutherland model can be obtained in the 2D YM theory with inserted Wilson line along the cylinder in the particular representation, which amounts to the interaction of the holonomy eigenvalues \cite{gn1,gn2}. The Calogero-Sutherland potential reads
\beq
V(\th_i-\th_j)= \nu^2 \sum_{i<j} \frac{1}{\sin^2(\th_i-\th_j)}
\eeq
The coupling $\nu=\beta(\beta-1)$ corresponds to the specific choice of the representation ($\nu$'s
power of the fundamental one). It is known that the eigenfunctions are Jack polynomials $J^\beta_{\lam}$:
\beq
\Psi\left(\mathbf{p} | \boldsymbol{\theta}  \right)=\frac{1}{\sqrt{N!}} J^\beta_R(e^{i
\th_1},\dots,e^{i \th_N}) \Delta^\beta(e^{i \boldsymbol{\th}})
\label{eq:jack}
\eeq
In (\ref{eq:jack}) we have introduced an auxiliary Young diagram $R$, with $n_i=p_i+i$. The energy,
$E$, is given by the deformed quadratic Casimir of $R$:
\beq
E(\mathbf{n})=\cfrac{1}{2m}\left( \cfrac{2 \pi}{L} \right)^2 \sum_i n_i(n_i- 2 i \beta + N + 1)
\eeq
The Calogero superpotential, according to (\ref{eq:super}), becomes:
\begin{equation}
W(\mathbf{\th})=\gamma \sum_{i<j} \log \sin(\th_i-\th_j), \qquad 2D \gamma(\gamma-1)=\beta(\beta-1)
\label{eq:calog-sup}
\end{equation}

Now we see that the partition function for the $\beta$-deformed YM on a cylinder \eq{eq:partition_cylinder} (where instead of Schur polynomials we used Jack polynomials) coincides with the transition amplitude for Calogero system. Boundary holonomies give boundary conditions for particles and the coupling constant, $g_C^2$, should be identified with the combination $t\left(\frac{2 \pi }{L} \right)^2$. The eigenstates of the Calogero model behave as anyons (see \cite{poly} for a review). So, various deformations of the 2D YM theory are described by free particles with generalized statistics. The partition function of $\beta$-deformed YM on $S^2$ corresponds to the reunion probability of particles interacting with the Calogero potential. The DK phase transition in this case happens at the critical time of the process $t_{cr}$, given by
\beq
\left(\cfrac{2 \pi}{L} \right)^2 \cfrac{t_{cr} N}{m} = \cfrac{\pi^2}{\nu}
\eeq

The Fokker-Planck equation in the dimensionless time, $s$, associated with the Jack process, reads
\beq
\frac{\partial P(s,\mathbf{\th})}{\partial s}=\sum_{i=1}^N \frac{\partial^2
P(s,\mathbf{\th})}{\partial x_i^2} + \sum_{i=1}^N \frac{\partial}{\partial
x_i}\left(\frac{\partial W(\mathbf{\th})}{\partial x_i}P(s,\mathbf{\th})\right)
\label{eq:fp2}
\eeq
Eq.(\ref{eq:fp2}) is the master equation for the "Dyson Brownian particles" with the Langevin
stochastic dynamics $\mathbf{x}(s)\equiv \{x_1(s),...,x_N(s)\}$ in a quantized stochastic time,
$s$, and in the potential (\ref{eq:calog-sup}):
\beq
\frac{dx_i(s)}{ds} = -\frac{\partial W(\mathbf{x})}{\partial x_i(s)} + \eta_i(s); \qquad
(i=1,...,N)
\label{eq:lan}
\eeq
where $\eta_i(s)$ is a uncorrelated white noise, $\left<\eta_i(s)\eta_j(s')\right>=
\delta(s-s')\delta(i-j)$. Writing $x_j$ in the form $x_j=e^{i\theta_j}$, and considering the
hydrodynamic limit $N\to\infty$, $s\to\infty$ with $\frac{N}{s}=u={\rm const}$, we end up with the
Langevin equation
\beq
\frac{d\theta_j(s)}{ds}=-\gamma \sum_{j\neq k} \cot(\theta_j(s)-\theta_k(s)) + \eta_j(s);
\qquad (i=1,...,N)
\label{eq:fp3}
\eeq

\section{DK transition and duality}

\subsection{From circle to line with harmonic trap: Fourth order Hamiltonian.}

So far we considered systems of fermions (or anyons) on the circle where the DK transition was interpreted as a kind of an orthogonality catastrophe for fixed initial and final states. Let us exploit the duality found in \cite{nekrasov97} which relates the trigonometric Calogero model on a circle of radius $R$ to the rational Calogero model on an infinite line in a harmonic potential well. Consider two Hamiltonians for the $N$-particle system: on a circle, $H_{I,2}$, and on a line with an additional parabolic well, $H_{II,2}$, where:
\beq
\begin{cases}
\disp H_{I,2}= \sum_{i=1}^N \frac{\partial_i^2}{\partial x_i^2} + \sum_{i\neq
j}^N\frac{g_{I}(g_{I}-1)}{\sin^2 ((x_i-x_j)/R)} & \mbox{for a circle}
\medskip \\
\disp H_{II,2}= \sum_{i=1}^N \frac{\partial_i^2}{\partial x_i^2} + \sum_{i\neq
j}^N\frac{g_{II}(g_{II}-1)}{(x_i -x_j)^2} + \omega^2 \sum_i x_i^2 & \mbox{for a harmonic trap on a line}
\end{cases}
\label{eq:nek}
\eeq
The coefficient, $\omega$, in the parabolic potential is related to the radius, $R$, of the circle as $\omega=\frac{1}{R}$. It was proved that the coupling constants
in two systems coincide, i.e. $g_{I}=g_{II}$, therefore if we choose, say, the coupling $g=1$ corresponding to the free fermions in one system,
we obtain the free fermions in the second system as well.

The correspondence between the Hamiltonians in two systems is as follows \cite{nekrasov97}
\beq
H_{II,2}= H_{I,1} = P_{I},\qquad H_{I,2}= H_{II,4}
\label{eq:nek2}
\eeq
where $H_{I,k}$, $H_{II,k}$ and $P_{I}$, $P_{II}$ are the $k$-th Hamiltonians for the systems $I$ and $II$. That is, time and coordinate directions in two systems are different and the mapping of the evolution operators reads as
\beq
U_I = \exp(it_{I,1}H_{I,1} + it_{I,2}H_{I,2}) \quad \leftrightarrow \quad
U_{II} = \exp(it_{II,2}H_{I,2} + it_{I,4}H_{I,4})
\eeq
For both nontrivial "times", $t_1$ and $t_2$ in the system $I$, we get the non-vanishing "times" $t_2$ and $t_4$ for quadratic and quartic Hamiltonians in the system $II$. It means that in the system $II$ we have the fermions perturbed by the Hamiltonian of the fourth order in coordinates and momenta.

Having in mind this duality we could ask how the DK transition for a system on a circle translates to the DK transition for a system on a line in a harmonic trap, and what is the meaning of the winding (natural for a circle geometry) for the dual system on a line? Naively there is no evident place for the winding in the harmonic potential on a line at all. For the system on a circle winding is defined as $N_I=\int \dot{q}dt_{I,2}$, hence in the dual system on a line we have to consider the evolution of some periodic variable with respect to $t_{II,4}$. Let us introduce the periodic (angular) variable on the phase plane. The "time", $t_{II,4}$, plays the role of the perturbation of the harmonic trap potential, therefore we have to deal with the flow in the parameter space. Fortunately these subtle points fit together to provide the natural candidate for a winding in the type II system.

Recall that in the Hamiltonian dynamics there is the so-called Hannay angle which describes the nontrivial bundle of the angle variable over the parameter space. The action-angle variables can acquire the nontrivial holonomy when moving in the parameter space. It is the semiclassical counterpart of the Berry connection known in the quantum case. Typically we have nontrivial Hannay connection when there are the selfintersection of the Lagrangian submanifolds in the phase space. This is exactly what we have here: the natural winding in the perturbed harmonic trap is
\beq
N_{II}=\int \dot{\alpha} dt_{II,4}
\eeq
where $\alpha$ is the angular variable on the phase space in the system $II$. We presented this argument for the single degree of freedom, however it seems to work for multi-particle case as well.

To suggest the interpretation of DK transition, consider first the case of one degree of freedom.
The condition that we start at fixed value of $q$ in system $I$ means that we consider the fixed angle in the system $II$, since for the case of fermions without interaction, the momentum and coordinate in system $I$ gets mapped precisely onto the action and angle in the system $II$. Now turn on $t_{II,4}$ and follow the trajectory of the point in the phase space as a function of  $t_{II,4}$. The "reunion" process in the system $I$ is translated to the process in the system $II$, where the trajectory of the point reach the same angle coordinate after evolution in the parameter space. It is clear that the trajectory can wind around the origin, yielding some phase. Certainly, in the quantum formulation, we have some distribution of winding numbers and the very issue can be thought as the investigation of the "orthogonality catastrophe" under the "evolution in the parameter space". To some extend, it can be thought of as an example of dynamics in the RG time.

We have illustrated above meanings of the winding number and of the reunion process in the system $II$ for the one-body problem where the DK transition is absent. In the large-$N$ limit, i.e. in the hydrodynamic approximation, the system of free fermions on the line is described by the forced Hopf equation \cite{abanov3}. There is the natural Hall-like droplet before turning on the $t_{II,4}$ perturbation. The droplet form evolves under the perturbation and we can pose the "reunion" problem for the droplet state.
The DK transition means that the sum of individual windings of fermions along the flow in the parameter space does not vanish in the strong coupling case. This picture resembles the emergence of the two-cut support in the spectral density in the matrix model discussed in related issue  \cite{gopakumar}.

\subsection{Vicious walks on a line in a harmonic trap: Ground state.}

In this subsection we formulate one more way to get the DK-like phase transition in the system of fermions in the harmonic trap. The straightforward view on vicious walks on a line is as follows. Consider the system of $N$ fermion worldlines (vicious walkers) on a line in a parabolic confining potential $U(x_1,...x_N)=\omega^2\sum_{j=1}^N x_j^2$. The trajectories start from the configuration
$\mathbf{x}(0)=(x_1(0),...,x_N(0))$ and arrive after time $t$ at the configuration $\mathbf{x}(t)=(x_1(t),...,x_N(t))$. We are interested in the reunion probability, $P(t,\mathbf{x}(0),\mathbf{x}(t))$, which can be explicitly written as
\beq
P(t,\mathbf{x}(0),\mathbf{x}(t)) = \sum_{\mathbf{n}}
\Psi_N(\mathbf{n}|\mathbf{x}(0))\Psi^*_N(\mathbf{n}|\mathbf{x}(t)) e^{-tE(\mathbf{n})}
\label{sw:01}
\eeq
where $\Psi_N(n|\mathbf{x}(t))$ is a Slater determinant
\beq
\Psi_N(\mathbf{n}|\mathbf{x}(t)) = \frac{1}{\sqrt{N!}}\det \psi(n_j|x_k(t))
\label{sw:02}
\eeq
and $\psi(n|x)$ is a single-particle solution of diffusion equation in a parabolic confining
potential:
\beq
-\frac{\partial^2}{\partial x^2} \psi(n|x) + \omega^2 x^2 \psi(n|x) = E_n \psi(n|x)
\label{sw:03}
\eeq
Solving (\ref{sw:03}), one gets up no normalization
\beq
\psi(n|x) \propto H_n(x \sqrt{\omega}) e^{-x^2 \omega^2/2}
\label{sw:04}
\eeq
where $H_n(x)$ are the Hermite orthogonal polynomials which satisfy the three-recurrence relation
\beq
H_{n+1}(y)=2yH_n(y)-2nH_{n-1}(y)
\label{sw:05}
\eeq
The explicit expressions of the eigenvalue, $E_{{\bf n}}$, and corresponding eigenfunction,
$\Psi_N(\mathbf{n}|\mathbf{x}(t))$, in (\ref{sw:02}), are
\beq
\left\{\begin{array}{l}
\disp E_{{\bf n}}\equiv E_{n_1,...n_N}=\sum\limits_{k=1}^N n_k \medskip \\
\disp \Psi_N(\mathbf{n}|\mathbf{x}(t)) \propto \frac{1}{\sqrt{N!}}\sum_{1\le n_1\le...\le n_N}^N \left|
\begin{array}{cccc} H_{n_1}(x_1) & H_{n_1}(x_2) & ... & H_{n_1}(x_N) \\ H_{n_2}(x_1) & H_{n_2}(x_2)
& ... & H_{n_2}(x_N) \\ \vdots & \vdots &  & \vdots \\ H_{n_N}(x_1) & H_{n_N}(x_2) & ... &
H_{n_N}(x_N) \end{array} \right| e^{-\frac{\omega^2}{2}\sum\limits_{n=1}^N x^2_n(t)}
\end{array} \right.
\label{sw:slat}
\eeq
Substituting (\ref{sw:slat}) into (\ref{sw:02}), we get the expression for the reunion probability,
which gets essentially simplified at $t\to\infty$ because in this limit only the ground state wavefunction with ${\bf n}_{gr}=(n_1=1,n_2=2,...,n_N=N)$ survives in the sum (\ref{sw:02}). Thus, at $t\to\infty$ one has for the reunion probability
\begin{multline}
\Psi_N(\mathbf{x}) \equiv \lim_{t\to\infty}\Psi_N(\mathbf{n}|\mathbf{x}(t)) \propto \Psi_N({\bf
n}_{gr}|\mathbf{x})\; \Psi_N^*({\bf n}_{gr}|\mathbf{x}(t)) \propto \left|\begin{array}{cccc} 1 & 1 & ...
& 1 \\ x_1 & x_2 & ... & x_N \\ \vdots & \vdots &  & \vdots \\ x^{N-1}_1 & x^{N-1}_2 & ... &
x^{N-1}_N
\end{array}\right|^2 e^{-\omega^2\sum\limits_{n=1}^N x^2_n} \\ = \prod_{k>j}
\big(x_k-x_j\big)^2 e^{-\omega^2\sum\limits_{n=1}^N x^2_n}
\label{sw:06}
\end{multline}

\begin{figure}[ht]
\epsfig{file=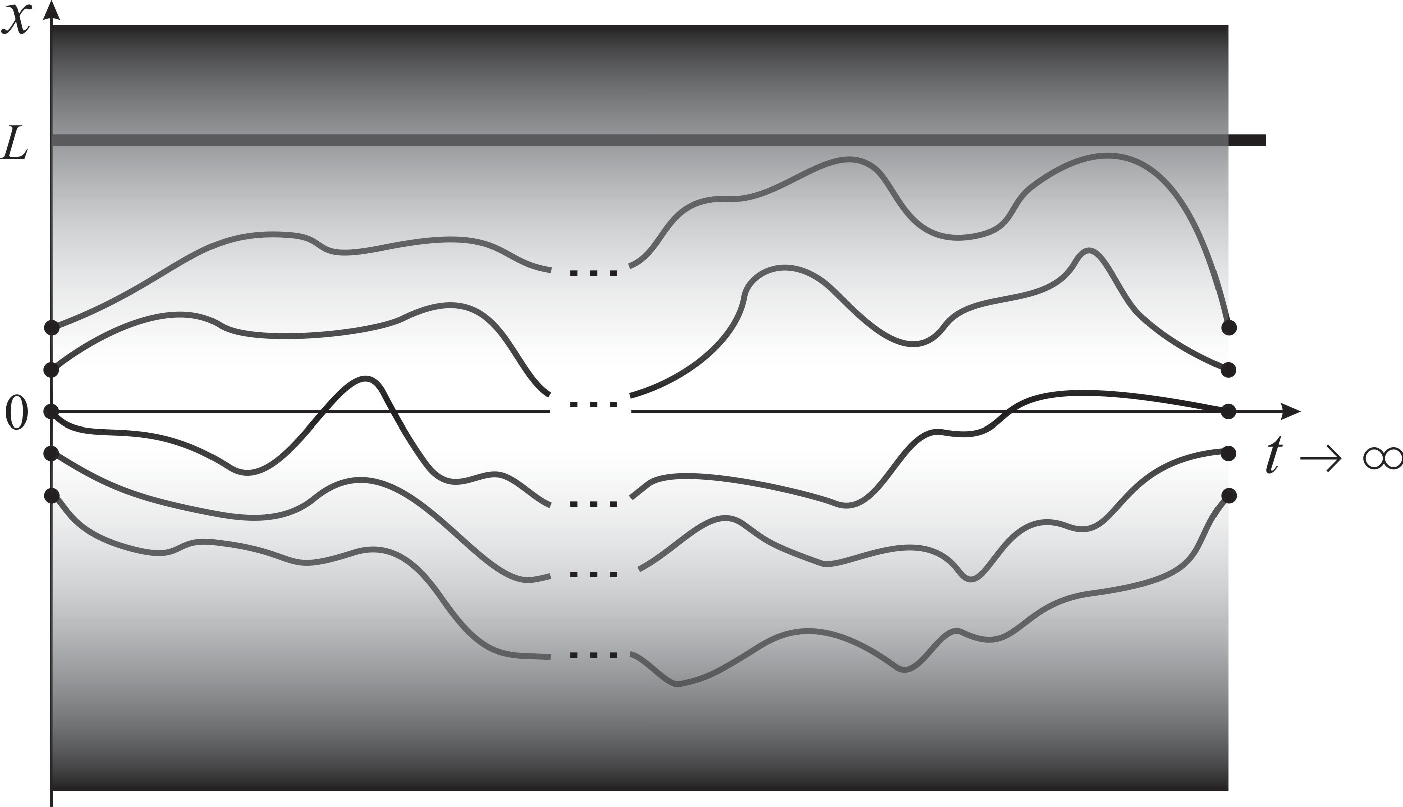, width=10cm}
\caption{The bunch of long fermion paths on an infinite support in a quadratic potential.
The topmost line is bounded from above at the distance $L$ from the origin.}
\label{fig:line}
\end{figure}

Now we can ask a question about the "conditional reunion probability", $\Psi_N(x_1<x_2<...<x_N<L)$,
where $L$ is the location of the upper boundary for the topmost fermionic path - see the
\fig{fig:line}. The bunch of $N$ fermion worldlines lives in the quadratic potential well,
$U(x_1,...x_N)=\omega^2\sum_{j=1}^N x_j^2$, schematically shown by the gradient color in the
\fig{fig:line}, which prevents the fermion paths to escape far from the region around $x=0$. So,
one can say that the fermion trajectories are "softly" supported from below, while "rigidly"
bounded from above.

The critical behavior in this system is formally identical to the one in the bunch of fermion
paths of finite length under the constraint $x_1<x_2<...<x_N<L$ on the position of the upmost line.
Comparing (\ref{sw:06}) and (\ref{eq:reunion2}), and taking into account that for VW ensemble the
transition occurs at the point given by (\ref{eq:crit}), we can conclude that the DK transition
point in harmonic potential on the line in the limit $t\to\infty$ is defined by the relation
\beq
\frac{\omega^2}{\pi^2} N =1
\label{eq:crit2}
\eeq

\section{Topological susceptibility -- the way to superfluidity}

\subsection{Winding distribution in a superfluid}

In this Section we conjecture that the instanton driven strong coupling phase enjoys a kind of the
superfluid component. The ground for this conjecture comes from the particular representation of
the density of the superfluid component \cite{pollock1,pollock2,prok} in terms
of the microscopic degrees of freedom. Namely, we will find that the microscopic winding number
$M$ has Gaussian distribution in both cases. We will start from reviewing this result in the theory of superfluidity.

Let us start from effective macroscopic low energy mean field theory described in terms of
the complex field $\Psi(x)$ which can be considered as the condensate wave function. Twist $\phi_0$ desctibes the boundary condition
along particular(say $x$) closed direction:
\beq
\Psi(L_x) = e^{i \phi_0} \Psi(0)
\eeq
$L_x$ is the size in this direction.

The response of the system to the twist serves a  tool to describe the superfluid component.
Introduce the topological susceptibility as the second derivative of the free-energy $F$:
\beq
\rho_s^{\phi} = \left.\frac{L_x^2}{\eta V}\frac{d^2F}{d^2 \phi_0}\right|_{\phi_0=0}
\eeq
where $L_x$ is the size of the system in the $x$-direction, $V$ - total volume and $\eta=\frac{\hbar}{m}$

A remarkable fact is that the partition function of the whole system is given by a sum over windings:
\beq
\label{z_phi}
Z(\phi_0) \propto  \sum_{M=-\infty}^{M=+\infty} \exp \left( -\frac{T L_x}{2\rho_s}M^2 -iM\phi_0 \right)
\eeq
In what follows we shall use the occurrence of the Gaussian distribution over the total microscopic
winding number $M$ in the instanton driven phase above the DK transition point as the indication of
the non-vanishing superfluid density.

It is remarkable that the topological susceptibility is closely related to the density of
superfluid component \cite{pollock1,pollock2}. In $d\ge 3$ $ \rho_s^{\phi} = \rho_s$ while in lower
dimensions they do not coincide however the relation holds \cite{ps}
\beq
\rho_s^{\phi} = \rho_s \left( 1 - \frac{4\pi^2 \eta \rho_s\left<I^2\right>}{TL_x^2} \right)
\label{density}
\eeq
where $I$ is the winding number for the superfluid component.	

Therefore the non-vanishing topological susceptibility implies the non-vanishing density of the
superfluid component. Let us emphasize that in the one-dimensional space the superfluidity is a
subtle phenomena and not all typical superfluid phenomena hold in $1+1$.  It was argued that the drug
force could substitute the order parameter \cite{caux} however the susceptibility as the measure of
the superfluid density still works.

\subsection{$\theta $-term versus twist in a superfluid}

Let us demonstrate a close analogy between the twist and the conventional $\theta$-term in the 2D
YM theory, which implies that the non-vanishing topological susceptibility in the 2D YM theory can be considered
as the counterpart of the superfluid density. The $\theta$-term can be added to the 2D YM action as
a chemical potential for the topological charges of the abelian instantons, which are classical
solutions to the equations of motion in the 2D YM on $S^2$. In the Hamiltonian framework the
$\theta$-term enters in the following way:
\beq
Z_g=\int [d\mathbf{A}] [d \mathbf{\Phi}] \exp \left(\int_{\Sigma_g} d\sigma dt \Tr\, \mathbf{\Phi} \mathbf{F}
+ \frac{g_{YM}^2}{N} \Tr\, \mathbf{\Phi}^2 +
\theta \Tr\, \mathbf{\Phi}\right).
\eeq
and the partition function  reads
\beq
Z_g(g_{YM},A,\theta)=\sum_{\boldsymbol{n}}\Delta^{2-2g}(\boldsymbol n) \exp\left(-\frac{Ag_{YM}^2}{2N}\sum
n_i^2 -i\theta \sum n_i\right)
\label{eq:Z_theta}
\eeq
where $\theta$ parameter plays the role of the chemical potential for the total $U(1)$ instanton
number.

What is the counterpart of the $\theta$  parameter for the vicious walkers? Comparison with the
instanton representation of YM partition function yields the following term in the Hamiltonian of
the particle system on a circle
\beq
\Delta H = \theta \frac{L}{2\pi t} \sum_i p_i
\eeq
Therefore the YM-VW mapping implies that $\theta$-term is the chemical potential for the total
winding in the random walk problem similar to the twist $\phi_0$ mentioned above.

Now we are at position to utilize the relation (\ref{density}) in the YM and VW frameworks. The key
point is that the distribution of  microscopic total winding number in RW or topological charge in
2D YM can be obtained explicitly \cite{lw}. For RW/YM at $t<t_c=L^2/4N$, the probability for a system to have
a non-zero winding is exponentially small,
\beq
\mathbb P(M=0)=1-\mathcal O(e^{-cN}).
\label{eq:prob_subcrit}
\eeq
Near the critical time, $t=t_c$, there are finite probabilities to get $M=0, M=1, M=-1$
\beq
\mathbb P(M=0)=1- \frac{q(s)}{2^{1/3}N^{1/3}} +...;  \qquad \mathbb P(M=1)= \mathbb P(M=-1)=
\frac{q(s)}{2^{4/3}N^{1/3}}+...
\label{eq:prob_crit}
\eeq
where $q(s)$ is the solution to the Painleve II equation
\beq
q''(s)= sq(s) +2 q(s)^3
\eeq
which has the Airy asymptotics
\beq
q(s)= Ai(s),\qquad s\rightarrow +\infty
\eeq
The variable $s$ is defined as follows
\beq
t=\pi^2 \left( 1 - \frac{s}{(2N)^{2/3}} \right)
\eeq
Let us emphasize that $\left<M^2\right>\neq 0$ at the critical point.

At $t>t_c$, the probability to have a specific total winding number $M$ is given by a Gaussian
distribution centered at zero,
\beq
\mathbb P(M)=Ce^{-\kappa M^2}+\mathcal O(N^{-1}).
\label{eq:prob_supercrit}
\eeq
where $C$ - is a mere normalization constans and $\kappa$ is an implicit function of $t$, given in terms of full elliptic integrals:
\beq
\kappa=-\frac{\pi K(k')}{K(k)}, \qquad  t=8E(k)K(k)-(1-k^2)K^2(k), \qquad k' =\sqrt{1-k^2}.
\label{eq:kappa}
\eeq
Hence, comparing with eq. (\ref{z_phi}) we can map the strong coupling phase with the superfluid density
\beq
\rho_s^{-1}=\frac{\pi K(k')}{K(k)}
\eeq
and the very DK transition acquires the interpretation as transition to the superfluid phase. Let
us emphasize that our consideration can be considered as the evidence in favor of the superfluid
interpretation of DK transition however more detailed study is required.

\section{DK transition for black hole partition function}

In this Section using the relation between the magnetically charged extremal BH and 2D $q$-YM
theory we will make a conjecture concerning the meaning of DK phase transition at BH side and the Browinian stochastic dynamics of the branes representing magnetic charges of BH.

\subsection{$Z_{BH}= Z_{2D ~qYM}$}

Consider the partition function of the extremal $\mathcal{N}=2$ black hole with the electric, $p_i$, and magnetic, $q_i$, charges. The extremal charged SUSY BH in four dimensions is created by $N$ magnetic $D4$--branes wrapped around the non-compact four manifold $C_4 = O(-p) \rightarrow CP^1$ in the internal non-compact CY space  $M= O(p-2) \times O(-p) \rightarrow CP^1$. The moduli of the extremal BH solution are fixed by their charges via attractor equations.

The key point is the identification of the BH partition function in the mixed representation with
the 2D $q$-YM on some Riemann surface. The terms in the action on $D4$ brane wrapped $C_4$ involve
the chemical potentials for the electric charges
\beq
\delta S= \frac{1}{2g_s}\int _{C_4} {\rm Tr} F\wedge F + \frac{\theta}{g_s} \int _{C_4} {\rm Tr}
F\wedge K
\eeq
where $K$ is the K\"ahler class of $CP^1$ inside $C_4$. The path integral with this action
turns out to coincide with the partition function of 2dqYM and reads as
\beq
Z_{BH}=Z^{qYM}= \sum_{q_1,q_2} \Omega (q_0,q_1,N) \exp\left(-\frac{4\pi^2}{g_s}q_0 - \frac{2\pi
\theta}{g_s}q_1\right)
\eeq
where $\Omega (q_0,q_1,N)$ can be identified as the degeneracy of $D0 \to D2 \to D4$ bound states
with charges $(q_0,q_1,N)$ correspondingly. The partition function involves the chemical potentials
for $D0$ and $D2$ branes
\beq
\phi_0= \frac{4\pi^2}{g_s} \qquad \phi_1= \frac{2\pi \theta}{g_s}
\eeq
The topological string coupling $g_s$ is related with the 2d YM coupling $g_{YM}$ via relation
\beq
g_{YM}^2= pg_s
\eeq

\subsection{DK for extremal magnetic black hole}

Let us comment qualitatively on the following question: "What happens with the magnetically charged BH at the transition critical line?". First consider the equation defining the critical line in the $q$-YM:
\beq
A_{crit} = p^2 \log \left( 1+ \tan^2 \left( \frac{\pi}{p} \right) \right), \qquad  p>2
\eeq
where the magnetic $N$ $D4$ branes are wrapped around $O(-p)\rightarrow CP^1$ in CY. The critical
point in pure YM is restored at $p\rightarrow \infty$ limit when $g_sp$ remains finite. We expect the set of phase transitions (as in the $q$-YM case) when different color orientations of "monopole" $D2$ branes representing instantons in the 2D YM and 't Hooft loops in 4D $SU(N)$ theory on $D4$ branes dominate.

In terms of the "monopole" $D2$ branes, the DK phase transition sounds as follows. Consider the BH
with large magnetic charge $Q_m=pN$ and vary the chemical potential for the $D0$ branes which is
the K\"ahler class of the base sphere. At some value of the chemical potential we fall into the
strong coupling phase driven by the t'Hooft monopole loops in  representation with some charge
vector.  From the observer at the BH the transition looks as sudden transition from the electrically neutral BH to the state with the Gaussian distribution of the charges corresponding
to the 't Hooft loops on $D4$ branes.

We can speculate that the DK transition in the extremal magnetic BH corresponds to the transition
discussed in \cite{bolognesi}. For the electrically charged BH there is so-called superconducting
phase transition (see \cite{hartnoll} for the review) which corresponds holographically to the
transition to the  superconducting state in the boundary theory. The physics in the bulk behind
this transition allows the following interpretation. At some value of the chemical potential it
becomes favorable for the BH to polarize the bulk, and the electric charge at the horizon gets
screened while the effective condensate of the charged scalar in the bulk emerges. In the magnetic
case, we expect a similar picture: let us start with the magnetically charged BH and vary the
parameters of the solution. At the critical point it becomes favorable to have monopole wall
instead of the extremal magnetic BH. Similarly, we have magnetic polarization of the bulk, the
horizon gets discharged, and a kind of the monopole condensate emerges in the bulk. The transition
in the flat space takes place at
\beq
v=cg_s
\eeq
where $v$ is VEV of the scalar. Therefore, we have some matching with the above conjecture. The value of the scalars from the vector multiplet are fixed by the attractor mechanism. On the other hand, in \cite{bolognesi} the scalars from nonabelian group are fixed at the transition point as well.

\subsection{BH and random walks}

Since we know the relation between the partition function of the $q$-YM and the Brownian reunion
probability, we can question about the stochastic random walk interpretation of the black hole entropy counting. Remind that the near horizon geometry of the BH is $AdS_2\times S^2$ and the entropy counting certainly is related to the $AdS_2$ part of the geometry. Another inspiring relation concerns the interplay between the BH partition function and $c=1$ string at selfdual radius \cite{ooguri}. The fermion representation of $c=1$ model is related with the fermion representation of the 2D YM theory on the sphere.

So far, the random walk interpretation of the BH entropy concerns the behavior of the long string
near the BH horizon at the almost Hagedorn temperature. The idea goes back to \cite{susskind} when
it was suggested that the BH entropy comes from the degeneracy of the states of a single long string wrapped the BH stretched horizon. In \cite{polchinski} dealing with the representation of the string gas, it was argued that the single long string behaves as the random walker, and the string with unit winding number dominates near the Hagedorn transition \cite{kru}. This winding mode corresponds to the thermal scalar which becomes tachionic at the Hagedorn temperature. The picture of the wound long string as the random walker has been generalized for more general geometries in \cite{zakharov}. In particular, the relation between the Hagedorn and Hawking temperatures plays the key role for establishing this correspondence for the non-extremal black holes \cite{zakharov}.

The discussed picture becomes more rich in our case for extremal magnetically charged BH. First of
all let us emphasize that so far we considered the Brownian $D4$ branes which undergo the
stochastic dynamics. Moreover, we have large $N$ number of Brownian branes. The Brownian $D4$
branes are extended in CY space and the stochasticy comes from the interaction with the strings.
Since we are looking for the extremal BH the Hawking temperature vanishes. Each Brownian brane
carries a huge multiplicity due to the bound states with $D2$ and $D0$ branes and additional
t'Hooft loops.

Since the mixed partition function of the BH $Z(N+i\phi_0)$ can be identified with the reunion probability of $N$ stochastic fermions on a circle for the time $T$, $P_{reun} (N,T)$, to get the Bekenstein entropy as the function of the magnetic and electric charges $S(N,Q_E)$ we have to perform the Laplace transform with respect to the electric chemical potential. On the other hand since we have identified the electric chemical potential with the time of reunion on the RW side we have to perform the Laplace transform with respect to the reunion time $T$
\beq
\int dT \exp(-ET) \left<N,x=0|e^{-\hat H T}|N,x=0\right>\propto Tr \left( \frac {1}{\hat H-E}
\right)
\eeq
Hence we get the resolvent for the system of $N$ non-interacting fermions and the energy $E$ plays
the role of the electric charge $Q_E$ at the BH side. The imaginary part of the resolvent indeed is
the density of states at the fixed "`electric charge $E$".

\section{Conclusion}

Our study provides the moderate step towards clarification the universal aspects behind the DK
phase transitions. The three systems were considered: large-$N$ perturbed topological gauge
theories, stochastic dynamics of large $N$ random walkers with and without interaction and entropy
of extremal 4D BH with large $N$ magnetic charge. Because of their diversity, the study provides the complimentary insights and intuition.  We tried to interpret in physical terms several seemingly different phenomena dealing with the DK-type transition, paying especial attention to the
construction of the unified picture.

At the random walk side we discussed the dependence of the DK phase transition on the
out-of-equilibrium boundary conditions and reformulate the problem as a kind of the orthogonality
catastrophe for the system of the free fermions. The generalization of the free walkers case to the
stochastic process with integrable interaction has been done which is equivalent to the
consideration of the matrix model with generic $\beta$-ensemble.

We have treated the thermal and quantum noise at the equal footing since in both cases the Langevin
and FP equations rule the game. However, the interpretation of the DK transition is very different
in two cases. In the thermal case we consider the process in the real time and the DK transition
admits an interpretation as one or another version of the orthogonality catastrophe for the fermion
gas in the simplest case. In the framework of the stochastic quantization we deal with the
stochastic or holographic time and the DK transition acquires the new a bit counter-intuitive
interpretation. It claims that some phenomena happens at the \emph{finite} holographic or
stochastic time. It means the RG has some nontrivial feature at the finite RG scale.

We believe that quantization of the generic system via the matrix model and topological strings
\cite{krefl,marinoq} provides proper language for these issues. Indeed in this approach the
quantization is performed within the $\beta$-ensemble of the large N matrix model which as we have
seen could have the DK-like transition. On the other hand the obstacle at the finite RG holographic
time is usually treated as the appearance of the BH horizon. However as we have mentioned it is
necessary to distinguish between the worldsheet horizon and the horizon of the BH background and it
is the worldsheet horizon which seems to be responsible for the DK transition. Anyway the
holographic interpretation of the DK transition certainly deserves the further study especially due
to relation with the multiple radial SLE stochastic process.

We made a conjecture about the nature of the DK transition for the extremal BH hole with large
magnetic charge. Certainly, the additional work is required to verify it. Potentially the question
seems to be very important since on the RW side we deal with the free fermions and one could ask
the question if the formation of the BH horizon for the magnetic BH can be described in simple
fermionic terms. Since the only phenomena which happens with free fermions, is the DK transition it
is natural to suspect that it has something to do with the horizon formation and elimination. The
possible relation with the conjectured appearance of the superfluid component, adds additional
flavor to the problem and probably some superfluid property of the BH horizon in the membrane
paradigm can be suspected.

The reader has certainly realized that the unifying theme for the different approaches and
viewpoints considered in the paper is the formation of a kind of horizon starting from the state in
out-of-equilibrium during the stochastic evolution. We believe that results of the paper and the
conjectures made can be of use for further-coming studies in this direction.

\begin{acknowledgments}
We are grateful to K. Bulycheva for the collaboration at the early stage of the work. We would like to thank A. Abanov, Victor Dotsenko, D. Kharzeev, N. Nekrasov, G. Oshanin, J. Policastro, N. Prokof'ev, R. Santachiara, G. Schehr, N. Sopenko, P. Wiegmann, and A. Zhitnitsky for the useful discussions and comments. The work of A.G. was supported by Basis Foundation fellowship and RFBR grant 19-02-00214; A.G. also thanks Kavli Institute for Theoretical Physics, Santa Barbara and Simons Center for Geometry and  Physics for support and hospitality. S.N. acknowledges the support of the Basis Foundation fellowship No.19-1-1-48-1.

\end{acknowledgments}

\appendix

\section{$q$-deformed 2D Yang-Mills on a sphere $S^2$}

Generalization of the 2D Yang-Mills theory onto the 2D $q$-Yang-Mills involves the additional parameter $q=e^{-g_s}$, where $g_s$ is the string coupling constant. The 2D $q$-YM theory on a sphere can be mapped onto the nondeformed 2D YM on a cylinder with specific boundary holonomies. The partition function of the $q$-deformed theory can be written as a sum over representations of the gauge group, as in the non-deformed case. The essential difference is that the dimension of the
representation should be replaced by the quantum dimension:
\beq
Z^{(q)}(g,g_s,A)= \sum_{R} (\dim_q\, R)^{2-2g} q^{\frac{Ag_{s}}{2N}C_2(R)}e^{i\theta |R|},
\label{eq:q_partition}
\eeq
where the sum runs again over the representations $R$, $C_1(R)$ and $C_2(R)$ are the first and the second Casimirs of the representation $R$, $\theta$ is its conjugate value of $C_1(R)$ in the grand canonical ensemble, and the quantum dimension is given in terms of the $q$-numbers
\beq
\dim_q\, R=\prod_{i>j}\frac{\left[n_i-n_j+j-i \right]_q}{\left[ j-i
\right]_q}=\prod_{i>j}\frac{\sinh \frac{g_s}{2}\left( n_i-n_j+j-i \right)}{\sinh
\frac{g_s}{2}\left(j-i \right)}
\label{eq:dim_q}
\eeq

The partition function of the $q$-Yang-Mills theory on the sphere ($g=0$) coincides with the
partition function of the non-deformed theory on the cylinder.  Recall that for the
quantum dimension there exists the following expression:
\beq
\dim_q\, R=\chi_R(q^{-\rho}),\quad \rho_i = \frac{1}{2}-i
\eeq
Thus, the partition function of $q$YM (\ref{eq:q_partition}) coincides with the partition function
on a cylinder (\ref{eq:partition_cylinder}) upon the identification \cite{deharo1,deharo2,deharo3}:
\beq
\theta_k=ikg_s, \qquad g_s^{-1}=g_{YM}^2.
\label{eq:q_cyl}
\eeq
Besides, we have to set particles equidistantly in the complex plane since the boundary holonomy is
$q^{-\rho}$. The phase transition for the $q$YM has richer structure compared to non-deformed
case.

\appendix

\end{document}